\documentclass[12pt]{article} 
\renewcommand\appendix{\par
  \setcounter{section}{0}
  \setcounter{subsection}{0}
  \setcounter{figure}{0}
  \setcounter{table}{0}
  \renewcommand\thesection{Appendix \Alph{section}}
  \renewcommand\thefigure{\Alph{section}\arabic{figure}}
  \renewcommand\thetable{\Alph{section}\arabic{table}}
}

\usepackage{amsmath}  
\usepackage{amsfonts} 
\usepackage{graphicx} 
\usepackage[T1]{fontenc}
\usepackage{bm}
\usepackage{appendix}
\usepackage{amssymb}

\def\theequation{\arabic{section}.\arabic{equation}}

\newcommand{\be}{ \begin{equation}}
\newcommand{\ee}{\end{equation}} 
\begin{document} 
\def\theequation{\arabic{section}.\arabic{equation}} 
\begin{titlepage} 
\title{Solving the Vialov equation of glaciology in terms of  
elementary functions} 
\author{Valerio Faraoni$^1$\\ \\ 
{\small $^1$~Department of Physics and Astronomy, Bishop's University}\\ 
{\small 2600 College Street, Sherbrooke, Qu\'ebec Canada 
J1M~1Z7}\\
{\small vfaraoni@ubishops.ca}
}
\date{} \maketitle 
\thispagestyle{empty} 
\vspace*{1truecm} 
\begin{abstract} 

Very few exact solutions are known for the non-linear 
Vialov ordinary differential equation describing the 
longitudinal profiles of alpine glaciers and ice caps under 
the assumption that the ice deforms according to Glen's
constitutive relationship. Using a 
simple, yet wide, class of models for the accumulation rate 
of ice and Chebysev's theorem on the integration of 
binomial differentials, many new exact solutions of the 
Vialov equations are obtained in terms of elementary 
functions.

 \end{abstract} 

\begin{center} 
PACS:92.40.vk, 92.40.Cy, 92.40.vv
\end{center}
\end{titlepage}


\section{Introduction}
\setcounter{equation}{0}
\label{sec:1}

A part of the mathematical modelling of alpine glaciers and 
polar ice sheets and ice caps is the description of their 
longitudinal profiles, which is based on non-linear 
differential equations.  The microphysics and the rheology 
of ice play a crucial role in determining the shape of 
glaciers. A good model for the response of glacier ice to 
stress is Glen's law relating the strain rate tensor 
$\dot{\epsilon}_{ij}$ to stresses in the ice (Glen~1955)
\be\label{Glen}
{\dot{\epsilon}}_{ij} = {\cal A} \, \sigma_\text{eff}^{n-1} s_{ij}
\ee
where $s_{ij}$ is the deviatoric stress tensor,  
\be
\sigma_\text{eff} =\sqrt{ \frac{1}{2} \mbox{Tr}\left( 
\hat{s}^2 \right)}
\ee  
is the effective stress, and ${\cal A}$ 
is a (temperature-dependent) constant (Paterson 1994; 
Cuffey and Paterson 2010; Hooke 2005; Greve and Blatter 
2009). The value $n=3$ is  
adopted for glacier flow in most theoretical and 
modelling work.

Let $x$ be a coordinate along the glacier bed in the 
direction of the ice flow. Assuming incompressible and 
isotropic ice, steady state, a flat bed 
(this means that the bed is a plane which, in general, has
 non-zero slope), 
and Glen's law, the longitudinal glacier profile (or 
local ice thickness) $h(x)$ 
obeys the Vialov ordinary differential equation (Vialov 
1958; Paterson 1994; Cuffey and Paterson 2010; Hooke 2005; 
Greve and Blatter 2009)
\be\label{Vialov}
x \, c(x) = \frac{2{\cal A} }{n+2} \left( \rho g h \left| 
\frac{dh}{dx} \right| \right)^{n} h^2 \,,
\ee
where $c(x)$ is the accumulation rate of ice, that is, 
the flux density of ice volume in the $z$-direction 
perpendicular to $x$, with the dimensions of a velocity. 
The absolute 
value in Eq.~(\ref{Vialov}) is introduced when one looks 
for solutions in the finite interval $x \in \left[0,L 
\right] $. If $x=0$ and $x=L $ denote the glacier summit 
and terminus, respectively, then the local surface slope 
$dh/dx$ is negative and its absolute value must be taken. 
If instead $x=0$ denotes the glacier terminus while $x=L$ 
is the summit, it is $dh/dx>0$. For ice caps and ice sheets, 
once a solution for the 
longitudinal profile of half of a glacier is found in 
$\left[ 0, L \right]$, it is extended to the interval 
$\left[ -L, L \right]$ (or to $\left[0, 2L \right]$, 
respectively) by reflection about the vertical 
line $x=0$ (or $x=L$, respectively) passing through the 
summit. A 
consequence of this procedure is that the surface profile 
$h(x)$ of an ice cap or ice sheet 
is not differentiable at the summit, where the left 
and right derivatives of $h$ are finite and opposite and, 
usually, also at the terminus where the slope $dh/dx$ and 
the basal stress $\tau_b= -\rho g h \, dh/dx$ diverge (here 
$\rho$ is the ice density and $g$ is the acceleration of 
gravity). This is, however, common procedure in the 
literature (Paterson 1994; Cuffey and Paterson 2010; Hooke 
2005; Greve and Blatter 2009). The non-linearity of the 
Vialov equation~(\ref{Vialov}) is a direct consequence of 
the non-linearity of Glen's law~(\ref{Glen}).

The formal solution of the Vialov equation~(\ref{Vialov}) 
can be expressed as the integral
\be \label{h(x)}
h(x)=\left\{ \mp \frac{ 2\left(n+1\right)}{ n\rho g} \left( 
\frac{n+2}{2{\cal A}} \right)^{1/n} 
\int dx \left[ x \, c(x) \right]^{1/n} 
\right\}^{\frac{n}{2(n+1)} } \equiv A  
\left[ V(x) \right]^{\frac{n}{2(n+1)} } \,, 
\ee
where the upper sign applies if the summit is at $x=0$ and 
$dh/dx<0$, while the lower sign applies if $x=L$ is the 
summit,
\be
A  \equiv  \left[ \frac{ 2\left(n+1\right)}{n\rho g} 
\left(\frac{n+2}{2{\cal A}} \right)^{1/n}\right]^{\frac{n}{2(n+1)}}  \,,
\ee
and the integral
\be
V(x)  \equiv \int dx \left[ x \, c(x) \right]^{1/n} \,.
\label{V}
\ee
is determined up to an arbitrary integration constant $D$. 
A function $c(x)$ modelling the accumulation rate of ice 
must be prescribed. Even for simple choices of $c(x)$, the 
integral~(\ref{V}) can rarely be computed in terms of 
elementary functions, which has led to stagnation 
in the literature on this subject, but a few analytic 
solutions of the Vialov equation are known\footnote{Other 
analytic profiles (Nye 1951a; Nye 1951b; Faraoni and Vokey 
2015) follow from the rather 
unrealistic assumption of perfecly plastic ice used in the 
early days of theoretical modelling and when the 
deformation of the ice is irrelevant.} (B\"o{\dh}vardsson 
1955; Vialov 1958; Weertman 1961; Paterson 1972; Bueler 
2003; Bueler et al. 2005). Solutions in $\left[ 0,L 
\right]$ with $x=0$ and $x=L$ denoting the position of the 
glacier summit and terminus, respectively, include:

\begin{itemize}

\item $c=$~const., which yields the {\em Vialov profile} 
((Vialov 1958), see also (Paterson~1994; Cuffey and 
Paterson 
2010; Hooke 2005; Greve and Blatter 2009))
\begin{eqnarray}
h(x) &=& H \left[ 1- \left( \frac{x}{L} 
\right)^{\frac{n+1}{n} } \right]^{\frac{n}{2(n+1)}} 
\,,\label{Vialovprofile1}\\
&&\nonumber\\
H &=& \left[\left( \frac{2}{\rho g} \right)^n 
\frac{c(n+2)}{2{\cal 
A}} \right]^{\frac{1}{2(n+1)} } \sqrt{L} \,.\label{Vialovprofile2}
\end{eqnarray}

\item If $c(x)$ is chosen as a step function, the {\em 
Weertman-Paterson profile} is obtained by matching two 
Vialov solutions (Weertman 1961; Paterson 1972).

\end{itemize}

Assuming instead $x=0$ at the glacier terminus and $x=L$ at 
the summit, one obtains the following solutions.

\begin{itemize}

\item The model $c(x)=c_m x^m$ is used in the literature, 
with the value $m=0$ believed to be appropriate for ice 
caps and $m=2$ for alpine glaciers. In scaling theory, 
according to the Buckingham Pi theorem (Buckingham 1914), 
the mass balance rate is supposed to scale as $l^m$, while 
the characteristic thickness $h$ of a glacier or ice cap is 
assumed to scale with its characteristic length $l$ as 
$h\sim l^s$. The exponents $s=\frac{m+n+1}{2(n+1)} $ and 
$s=\frac{m+1}{n+2} $ are predicted by scaling theory for 
ice caps and for alpine glaciers, respectively (Bahr et 
al. 2015). The power law
\be\label{powerlaw}
h(x) =h_0 \, x^{\frac{n+m+1}{2(n+1)}} 
\ee
(with $h_0$ a constant) solves the Vialov 
equation~(\ref{Vialov}) with $c(x)=c_m x^m$. The exponent 
$\frac{n+m+1}{2(n+1)}$ was deduced in scaling theory 
(Bahr et al. 2015; Faraoni 2016). This solution includes 
the case $c=$~const. 
and also the profile $h(x)=h_0 \, \sqrt{x}$ which 
reproduces the parabolic profile first obtained by Nye 
under 
the simplifying assumption of perfectly plastic ice (Nye 
1951a; Nye 1951b), which is very different from the more 
realistic Glen law~(\ref{Glen}) but is obtained as the 
limit $n\rightarrow +\infty$ of Eq.~~(\ref{powerlaw}). As 
shown in Sec.~\ref{sec:2.1}, the 
profile $h(x)=h_0 \, \sqrt{x}$ is not restricted to the 
unrealistic assumption of perfectly plastic ice but is also 
a solution of the Vialov equation following from the 
realistic Glen law. This fact is significant because this 
profile is currently used 
in a number of applications ({\em e.g.}, (Benn and Hulton 
2010; Ng et al. 2010)) and is appropriate when the internal 
deformation of the ice is irrelevant.

\end{itemize}

In Sec.~\ref{sec:2} a simple, yet broad, model of the 
function $c(x)$ describing the ice accumulation rate is 
postulated and the Chebysev theorem on the integration of 
differential binomials is applied to the search of exact 
solutions of the Vialov equation in terms of elementary 
functions, in the form~(\ref{h(x)}). Infinitely many new 
solutions in terms of elementary functions can be obtained, 
some of which are reported in appendix~\ref{appendix:A}, 
while known solutions are re-derived. Sec.~\ref{sec:3} 
contains a discussion of these solutions and of the method 
employed.

\section{Chebysev theorem and Vialov equation}
\label{sec:2}

Let us return to Eq.~(\ref{Vialov}) and let us search for 
solutions of the form (\ref{h(x)}) when the 
integral~(\ref{V}) can be expressed in terms of elementary 
functions. A wide class of reasonable models for the 
accumulation rate function is the choice
\be\label{c(x)}
c(x)= a+b \, x^r \,,
\ee
where $a,b$, and $r$ are constants and where $r$ is chosen 
to be rational for reasons explained below. Special cases 
include:

\begin{enumerate}

\item $a=0$, $b>0$, $r>0$. In this case $x=0$ is the 
location of the glacier terminus corresponding to zero 
accumulation rate, while the glacier summit is at $x=L$, 
where the accumulation rate of ice assumes its largest 
value $c_\text{max}$.  Then it follows that $ 
b=c_\text{max}/L^r $. The choice $r=2$ is appropriate to 
describe alpine glaciers (Bahr et al. 2015). Although it is 
not done in the literature, a better model 
would assume $a<0$ to describe ablation at the glacier 
terminus.

\item $c(x)=a-|b|x^r$ with $a>0, b<0$, and $r>0$. In this 
case it is appropriate to locate the summit at $x=0$ and 
the terminus at $x=L$, with $c(x)$ a decreasing function of 
$x$ in $\left[0,L \right]$ vanishing at $x=L$ and with the 
constants assuming the values $a=c_\text{max}$, 
$b=-c_\text{max}/L^r$. An alternative choice consists of 
having $c(L)<0$ in order to describe ablation at the 
terminus.

\end{enumerate}

\noindent With the choice $r\in \mathbb{Q}$, the integral 
$V(x)$ falls into the category 
\be \label{I}
I\left(x; p, q, r \right)= \int dx \, x^p \left( a+b \, x^r 
\right)^q     \,, \;\;\;\;\;\;\;\;
p,q,r \in \mathbb{Q} \,, r\neq 0 
\ee
(if $r=0$ the integral is trivial). In practice, for 
glacier flow it is $p=q=1/n=1/3 \in \mathbb{Q}$. The 
integral~(\ref{I}) can be expressed in terms of an 
hypergeometric function,
\begin{eqnarray}
&&\int dx \, x^{1/3} \left( a+b \,x^r\right)^{1/3} 
=\frac{ 3x^{4/3} }{ 4\left( 4+r\right) \left( a+b \, 
x^r\right)^{2/3} } \nonumber\\
&&\nonumber\\
& &  \cdot \left[ ar\left( \frac{bx^r}{a} +1 
\right)^{2/3} {}_2F_1\left( \frac{2}{3}, \frac{4}{3r}; 
1+\frac{4}{3r}; \frac{-bx^r}{a} \right)+ 4\left( a+ b \, 
x^r\right) \right]  \,, 
\end{eqnarray} 
but this representation is of little 
use for practical purposes, for example when, in 
statistics,  
one needs a simple model of longitudinal glacier profile  
$h(x)$ to fit a large number of glaciers. For numerical 
studies of a single glacier, it is convenient to integrate 
numerically Eq.~(\ref{Vialov}) but for other problems a 
simple analytic formula for $h(x)$ is required. A 
necessary and sufficient condition for the 
integral~(\ref{I}) to be expressed in terms of elementary 
functions is the\\\\
\noindent {\bf Chebysev theorem} (Chebysev 1853; 
Marchisotto and Zakeri 1994):\\\\
{\em the integral~(\ref{I}) admits a representation in 
terms of elementary functions if and only if at least one 
of} 
$$ 
\frac{p+1}{r} \,, \;\;\;\; q \,, \;\;\;\; \frac{p+1}{r} + q 
$$
{\em is an integer.}

Since $n=3$, it is $p=q=1/3 \in \mathbb{Q} $, and $r$ in 
Eq.~(\ref{c(x)}) is chosen to be rational (in the 
glaciological literature $r$ is usually the integer~0 or~2). 
Atmospheric models which could provide hints to fix  
the function $c(x)$ are not currently coded to 
have the ability to discriminate 
between a real number $r$ and a rational approximation of 
it. One then has
\begin{eqnarray}
\frac{p+1}{r} &=& \frac{n+1}{nr}=\frac{4}{3r} \,,\\
&&\nonumber\\
\frac{p+1}{r} + q &=& \frac{n+1+r}{nr}=\frac{4+r}{3r}\,.
\end{eqnarray}
Given the freedom in the choice of the parameters $a,b$, 
and $r$ of the model~(\ref{c(x)}), one requires that $r\in 
\mathbb{Q}$ and searches for values of $r$ such that 
$(p+1)/r$ or $q+(p+1)/r$ are integers.  

\begin{itemize}

\item By imposing that $\frac{4}{3r} \equiv m_0 \in \mathbb{Z}$, one obtains 
$r \equiv \frac{4}{3m_0}$, ~$m_0=1,2,3, \, ... \,, +\infty $. This 
choice produces the sequence of values of $r$
\begin{eqnarray}
&& \frac{4}{3} \simeq 1.33, \;\; \frac{2}{3} \simeq 
0.667,\;\; 
\frac{4}{9} \simeq 0.444,\;\; \frac{1}{3} \simeq 0.333,\;\; 
\frac{4}{15} \simeq 0.267,\;\; \frac{2}{9} \simeq 
0.222 , \nonumber\\
&&\nonumber\\
&& \frac{4}{21} \simeq 0.190, \;\; \frac{1}{6} \simeq 0.167,\;\;
\frac{4}{27} \simeq 0.148,\;\; \frac{2}{15} \simeq 
0.133,\;\; 
... \;\; \,,  0 \,. \label{m1} 
\end{eqnarray}

\item Imposing $ \frac{4+r}{3r}  \equiv m_0 \in \mathbb{Z}$ gives 
$r=\frac{4}{3m_0-1}$, ~$m_0=1,2,3, \, ... \, , +\infty$ and the sequence 
of values of $r$
\begin{eqnarray}
&& 2 ,\;\; \frac{4}{5} = 0.8,\;\; 
\frac{1}{2} = 0.5,\;\; \frac{4}{11} \simeq 0.364,\;\; 
\frac{2}{7} \simeq 0.286,\;\; \frac{4}{17} \simeq 
0.235,\;\; 
\frac{1}{5} = 0.2,\nonumber\\
&&\nonumber\\
&&  \frac{4}{23} \simeq 0.174,\;\; 
\frac{4}{23} \simeq 0.174,\;\; \frac{2}{13} \simeq 
0.154,\;\;  ... \;\; \,, 0 \,.\label{m2} 
\end{eqnarray}
\end{itemize}

Not all these values of $r$ are appropriate from the 
glaciological point of view to describe the 
ice accumulation rate~(\ref{c(x)}). However, the values~0 and~2 
universally used 
in the literature, and many values of potential interest 
lying between these two extremes, are reproduced. The value 
$r=0$ is usually suggested for ice caps and ice sheets 
while the value $r=2$ is suggested for alpine glaciers 
(B\"o{\dh}vardsson 1955; Vialov 1958; Weertman 1961; 
Paterson 1972; Bueler 2003; Bueler et al. 2005).  The 
representation of the integral $V (x)$ in terms of 
elementary functions falling into the range covered by the 
Chebysev theorem include the following special cases.

\subsection{Choice $c(x)=$~constant}
\label{sec:2.1}

The choice $c(x)=$~constant can be obtained by setting 
$b=0$ (in which case $r$ drops out of the discussion) or 
when $b\neq 0$ with $r=0$ (in which case the Chebysev theorem 
as stated does not apply). In both cases the integration is trivial 
and, in the first case, one obtains 
\be
V(x)=\frac{na^{1/n}}{n+1} \, x^{1+1/n}+ D \,,
\ee
where $D$ is an integration constant, and the 
longitudinal glacier profile 
\be
h(x)=A \left[  V(x) \right]^{ \frac{n}{2(n+1)} } = A \left[ 
\frac{na^{1/n} }{n+1} \, x^{1+1/n}+D \right]^{ 
\frac{n}{2(n+1)} } \,,
\ee
which reproduces the Vialov 
profile~(\ref{Vialovprofile1}), (\ref{Vialovprofile2}) 
always associated with the choice $c=$~const. 
in the glaciological literature (Paterson~1994; Cuffey and 
Paterson 2010;  
Hooke 2005;   Greve and Blatter 2009).
Setting $D=0$ yields the parabolic profile $h(x)=h_0 
\, \sqrt{x}$ irrespective of the value of $n$.

\subsection{Choice $c(x)=b \, x^r$}

In this case, with $a=0$ and $b>0$, and without choosing a 
specific value of $r$, one obtains the 
integral\footnote{Strictly speaking, in this degenerate 
case there is no need to assume that $r\in \mathbb{Q}$ 
and  
use the Chebysev theorem. In fact, these ingredients were 
not assumed in the recent work 
(Bahr et al. 2015) deriving this power law solution.}
\be
V(x)=\frac{3b^{1/3}}{4+r} \, x^{(4+r)/3} +D \,.
\ee
The corresponding  longitudinal glacier profile is 
\be
h(x) = A \left[ \frac{3b^{1/3}}{4+r} \, x^{(4+r)/3} 
+D \right]^{3/8} \,.
\ee
Setting $r=0, D\neq 0$ reproduces $c=$~const. and gives the 
Vialov 
profile~(\ref{Vialovprofile1}) and~(\ref{Vialovprofile2}).  
Setting instead the integration constant $D$ to zero yields 
$h(x)=h_0 \, x^{(4+r)/8} $. As already noted, the value $r=2$ 
is appropriate to describe alpine glaciers (Bahr et al. 
2015; Faraoni 2016) and gives $h(x) \propto 
x^{3/4}$. Setting instead $r=0$, which is appropriate for 
ice caps, yields the well known profile $h(x) \propto 
\sqrt{x}$ (Paterson~1994; Bahr et al. 2015). 

The choice $c(x)=b\, x^r$, usually written as $c(x)=c_m 
x^m$, reproduces the power law solution $h(x) \propto x^{ 
\frac{n+m+1}{2(n+1)}} $ of (Bahr et al. 2015; 
Faraoni 2016). In fact, setting 
$r=m$ and $n=3$ yields $h \sim x^{(4+r)/8}$.

As $x$ becomes large the highest order 
term is dominant and, in all of these solutions, the 
profile then approaches $ h(x) \sim \sqrt{x}$.

Other examples of 
plausible models of the accumulation rate 
$c(x)=a+b \, x^r$ leading to representations of the 
integral~(\ref{V}) in terms of elementary functions and to 
relatively simple exact profiles are reported in 
appendix~\ref{appendix:A}.

\section{Discussion}
\label{sec:3}

Analytic expressions describing longitudinal glacier 
profiles are needed in several problems of glaciology ({\em 
e.g.}, (Thorp 1991; Ng et al. 2010;  Benn and Hulton 
2010)). However, under the realistic and well tested 
assumption that glacier ice deforms according to Glen's
constitutive relationship~(\ref{Glen}), the Vialov ordinary 
differential equation~(\ref{Vialov}) ruling 
these longitudinal glacier profiles is non-linear and 
obtaining analytic solutions in closed form in terms of 
elementary functions is difficult. Only a few exact 
solutions are known in the literature (B\"o{\dh}vardsson 
1955; Vialov 1958; 
Weertman 1961; Paterson 1972; Bueler 2003; Bueler et 
al. 2005). By assuming a simple, 
yet general, model for the accumulation rate of ice 
appearing in 
Eq.~(\ref{Vialov}), the Chebysev theorem provides a 
necessary and sufficient condition for the 
integral~(\ref{V}) expressing a formal solution of the 
Vialov equation to be represented in terms of elementary 
functions. The solutions provided by the Chebysev theorem 
include the known solutions, with the exception of the 
B\"o{\dh}vardsson, Vialov, and Bueler profiles 
(B\"o{\dh}vardsson 1955; Vialov 1958; Bueler 2003; Bueler 
et al. 2005).

The initial condition $h=0$ of the Vialov 
equation~(\ref{Vialov}) is imposed at the glacier terminus 
($x=0$ or $x=L$, depending on the geometry adopted, which 
determines also the sign of $dh/dx$), which is a singular 
point of the equation corresponding to divergent surface 
slope $dh/dx$. In this situation, the usual uniqueness 
theorems for ordinary differential equations ({\em e.g.}, 
(Brauer and Noel 1986)) do not hold and this is the reason 
why one can find multiple solutions of the Vialov equation, 
and why the solutions obtained by using the Chebysev 
theorem do not always generate the well known Vialov (1958) 
profile, and do not reproduce other profiles 
(B\"o{\dh}vardsson 1955;  Bueler 2003; Bueler et al. 2005).

An infinite number of solutions in terms of elementary 
functions is guaranteed by the Chebysev theorem, 
corresponding to rational values of the constant $r$, and 
they can be found easily with computer algebra. The current 
models for the ice accumulation rate $c(x)$ are very 
unsophisticated ($c=$~const. being perhaps the most popular 
choice) and the 3-parameter choice $c(x)=a+b \, x^r$, $r\in 
\mathbb{Q}$ allows freedom to extend these models. Of 
course, other functional choices may be appropriate to 
model the accumulation rate $c(x)$ and, at the same time, 
provide analytic profiles $h(x)$. However, exact solutions 
of the Vialov equation~(\ref{Vialov}) in simple form have 
been hard to 
find and sometimes they correspond to unintuitive choices 
of $c(x)$ which make the corresponding analytic profile 
$h(x)$ more of a toy model achieving one desired physical 
property than a realistic description of the shape of 
alpine glaciers and ice caps. This is the case of the 
Bueler profile (Bueler 2003; Bueler et al. 2005; 
Greve and Blatter 2009), which exhibits a finite basal 
stress $\tau_b=-\rho g h \, dh/dx$ at the glacier terminus, 
contrary to the Vialov and other profiles. The old 
Chebysev (1853) theorem extends the scope of existing 
analyses. The values of the parameters $a,b$, and $r$ in 
Eq.~(\ref{c(x)}) appropriate to particular geographic 
locations have to be determined by data-fitting and are 
expected to be different for different situations (alpine 
glaciers, polar ice caps, cirque glaciers, {\em etc.}).

\section*{Acknowledgments} 
This work is supported by Bishop's University.



\newpage

\newpage
\appendix
\section*{Appendices}
\renewcommand{\thesubsection}{\Alph{subsection}}

\subsection{Exact solutions of the Vialov equation for 
some rational values of $r$}
\label{appendix:A}
\def\theequation{A.\arabic{equation}}\setcounter{equation}{0}

In the range of parameters $\left( p, q, r \right)$ in 
which the Chebysev theorem is satisfied, computer algebra 
easily provides the integral~(\ref{V}) for the 
choice~(\ref{c(x)}) of $c(x)$.  Here some of these 
integrals and the corresponding longitudinal glacier 
profiles are reported for various values of the 
parameter $r$ listed in Eq.~(\ref{c(x)}), which correspond to values 
of  $m_0$ 
reported in Eq.~(\ref{m1}).
\begin{eqnarray}
r&=&\frac{4}{3}  \,, \\
&&\nonumber\\
V(x) &=& \frac{9\left( a+b \,x^{4/3}\right)^{4/3}}{16b} +D \,,\\
&&\nonumber\\
h(x) &=& h_0 \left[ \left( a+b \,x^{4/3}\right)^{4/3}+D\right]^{3/8} \,,
\end{eqnarray}
where $D$ is, as usual, an integration constant. For $D=0$ one 
obtains 
\be
h(x) = h_0  \left( a+b \,x^{4/3}\right)^{3/8} 
\ee
and, if also $a=0$, one obtains again $h(x)=h_0 \sqrt{x}$. 
Another possibility is
\begin{eqnarray}
r&=&\frac{2}{3} \,, \\
&&\nonumber\\
V(x) &=& \frac{9\left( a+b \, x^{2/3}\right)^{1/3} \left( 
-3a^2+abx^{2/3}+4b^2 x^{4/3} \right) }{56b^2} +D \,,\\
&&\nonumber\\
h(x) &=& h_0 \left[ D_0 +\left( a+b \, x^{2/3}\right)^{1/3} 
\left( 
-3a^2+abx^{2/3}+4b^2 x^{4/3} \right) \right]^{3/8}\,,
\end{eqnarray}
where $D_0$ is another constant. Other possibilities are:
\begin{eqnarray}
r&=&\frac{4}{9} \,, \\
&&\nonumber\\
V(x) &=& \frac{27}{560b^3} \left( 
a+b \, x^{4/9}\right)^{1/3} \left( 9a^3 -3 a^2 b x^{4/9} +2 
a b^2 x^{8/9} +14 b^3  x^{4/3} \right) +D  \,,\\
&&\nonumber\\
h(x) &=& h_0 \left[ D_0 + \left( 
a+b \, x^{4/9}\right)^{1/3} \left( 9a^3 -3 a^2 b x^{4/9} +2 
a b^2 x^{8/9} +14 b^3  x^{4/3} \right)\right]^{3/8} 
\end{eqnarray}
\begin{eqnarray}
r&=& \frac{1}{3} \,, \\
&&\nonumber\\
V(x) &=& \frac{9}{1820b^4} \left( 
a+b \, x^{1/3}\right)^{1/3} \left( -81a^4 +27 a^3 b x^{1/3} -18 
a^2 b^2 x^{2/3} +14 ab^3 x +140 x^{4/3} \right) \nonumber\\
&&\nonumber\\ 
&\, & +D \,,\\
&&\\
h(x) &=& h_0 \left[ D_0 + \left( 
a+b \, x^{1/3}\right)^{1/3} \right. \nonumber\\
&&\nonumber\\
&\, & \left. \cdot 
\left(  -81a^4 +27 a^3 b x^{1/3} -18 
a^2 b^2 x^{2/3} +14 ab^3 x +140 x^{4/3} \right) \right]^{3/8} 
\,;\nonumber\\
&&
\end{eqnarray}
\begin{eqnarray}
r&=&\frac{4}{15} \,, \\
&&\nonumber\\
V(x) &=& \frac{9}{5824 b^5} \left( 
a+b \, x^{4/15}\right)^{1/3} \left( 243a^5 -81a^4 b x^{4/15}    
+54  a^3 b^2 x^{8/15} -42 a^2 b^3 x^{4/5} 
\right. \nonumber\\
&&\nonumber\\
&\, & \left. +35 a b^4 
x^{16/15} +455 b^5 x^{4/3}  \right)  +D \,,\\
&&\nonumber\\
h(x) &=&  h_0 \left[ D_0+  
\left( 
a+b \, x^{4/15}\right)^{1/3} \left( 243a^5 -81a^4 b x^{4/15}    
+54  a^3 b^2 x^{8/15} -42 a^2 b^3 x^{4/5} 
\right. \right. \nonumber\\
&&\nonumber\\
&\, & \left. \left. +35 a b^4 
x^{16/15} +455 b^5 x^{4/3}  \right) \right]^{3/8} \,;
\end{eqnarray}
\begin{eqnarray}
r&=&\frac{2}{9} \,, \\
&&\nonumber\\
V(x) &=& \frac{27}{55328b^6} \left( 
a+b \, x^{2/9}\right)^{1/3} \left( -729a^6 +243 a^5 b x^{2/9} 
-162 a^4 b^2 x^{4/9} +126 a^3b^3  x^{2/3} 
\right. \nonumber\\
&&\nonumber\\
&\, & \left.  -105 a^2b^4 
x^{8/9} +91ab^5 x^{10/9} +1456 b^6x^{4/3}  \right) +D  
\,,\\
&&\nonumber\\
h(x) &=&  h_0 \left[ D_0 + \left( 
a+b \, x^{2/9}\right)^{1/3} \left( -729a^6 +243 a^5 b x^{2/9} 
-162 a^4 b^2 x^{4/9} +126 a^3b^3  x^{2/3} 
\right.\right. \nonumber\\
&&\nonumber\\
&\, & \left. \left.  -105 a^2b^4 
x^{8/9} +91ab^5 x^{10/9} +1456 b^6x^{4/3}  \right) \right]^{3/8} 
\,,
\end{eqnarray}
\begin{eqnarray}
r&=&\frac{4}{21} \,, \\
&&\nonumber\\
V(x) &=& \frac{9}{173888 b^7} \left( 
a+b \, x^{4/21}\right)^{1/3} \left( 6561 a^7 -2187 a^6 b 
x^{4/21}    
+1458  a^5 b^2 x^{8/21} -1134 a^4 b^3 x^{4/7} 
\right. \nonumber\\
&&\nonumber\\
&\, & \left. +945 a^3 b^4 
x^{16/21} -819 a^2 b^5 x^{20/21} +728 ab^6 x^{8/7} +13832 
b^7 x^{4/3}  \right)  +D \,,\\
&&\nonumber\\
h(x) &=& h_0 \left[ D_0+ \left( 
a+b \, x^{4/21}\right)^{1/3} \left( 6561 a^7 -2187 a^6 b 
x^{4/21}    
+1458  a^5 b^2 x^{8/21} -1134 a^4 b^3 x^{4/7} 
\right.\right. \nonumber\\
&&\nonumber\\
&\, & \left. \left. +945 a^3 b^4 
x^{16/21} -819 a^2 b^5 x^{20/21} +728 ab^6 x^{8/7} +13832 
b^7 x^{4/3}  \right) \right]^{3/8} \,,
\end{eqnarray}
\begin{eqnarray}
r&=&\frac{1}{6} \,, \\
&&\nonumber\\
V(x) &=& \frac{9}{543400 b^8} \left( 
a+b \, x^{1/6}\right)^{1/3} \left( -19683 a^8 +6561 a^7 b 
x^{1/6}    -4374 a^6 b^2 x^{1/3} +3402 a^5 b^3 
\sqrt{x} \right.\nonumber\\
&&\nonumber\\
&\, & \left.  -2835 a^4 b^4 x^{2/3} +2457 a^3 b^5 x^{5/6} 
-2184 a^2 b^6 x +1976 a b^7 x^{7/6} +43472 b^8 x^{4/3}  
\right) +D  \,,\nonumber\\
&&\\
h(x) &=& h_0 \left[ D_0+ \left( 
a+b \, x^{1/6}\right)^{1/3} \left( -19683 a^8 +6561 a^7 b 
x^{1/6}    -4374 a^6 b^2 x^{1/3} +3402 a^5 b^3 
\sqrt{x} \right.\right.\nonumber\\
&&\nonumber\\
&\, & \left.\left.  -2835 a^4 b^4 x^{2/3} +2457 a^3 b^5 x^{5/6} 
-2184 a^2 b^6 x +1976 a b^7 x^{7/6} +43472 b^8 x^{4/3}  
\right) \right]^{3/8}  \,.\nonumber\\
&&
\end{eqnarray}


\begin{thebibliography}{99}

\bibitem{Bahrreview} Bahr DB, Pfeffer WT, Kaser G (2015) 
A review of volume-area scaling of 
glaciers. Rev. Geophys. 10:95-140.   
doi:~10.1002/2014RG000470

\bibitem{BennHulton} Benn DI, Hulton NRJ (2010) An excel spreadsheet 
program for reconstructing the surface profile of former mountain glaciers 
and ice caps. Comput. Geosci. 36:605-610.    
doi:~10.1016/j.cageo.2009.09.016

\bibitem{Bodvardsson} B\"o{\dh}vardsson G (1955) On the 
flow of ice-sheets and glaciers. J\"okull 5:1-8 

\bibitem{BrauerNoel} Brauer F, Noel JA (1986) {\em Introduction to 
Differential Equations With Applications}. Harper and Row, 
New York

\bibitem{Buckingham} Buckingham E (1914) On physically 
similar systems: illustrations of the use of dimensional 
equations. Phys. Rev. 4:435-376

\bibitem{Bueler1} Bueler E (2003) Construction of steady 
state solutions for isothermal shallow ice sheets, 
Fairbanks, AK, University of Alaska Fairbanks, Department 
of Mathematics and Statistics (UAF DMS Tech. Rep. 03-02)

\bibitem{Bueler2} Bueler E, Lingle CS, Kallen-Brown JA,  
Covey DN, Bowman LN (2005) Exact solutions and verification 
of numerical models for isothermal ice sheets. J. Glaciol. 
51:291-306

\bibitem{Chebysev} Chebyshev PL (1853) Sur l'integration 
des diff\'erentielles irrationnelles. Journal de 
mathematiques (series 1) 18:87-111

\bibitem{CuffeyPaterson} Cuffey KM, Paterson WSB (2010) 
{\em The Physics of Glaciers}. Elsevier, Amsterdam

\bibitem{FaraoniJOG} Faraoni V (2016) Volume/area scaling 
of glaciers and ice caps and their longitudinal profiles.
J. Glaciol. doi:~10.1017/jog.2016.79

\bibitem{FaraoniVokey2015} Faraoni V, Vokey MW (2015) 
The thickness of glaciers. Eur. J. Phys. 36:055031/1-11. 
doi:10.1088/0143-0807/36/5/055031

\bibitem{Glen} Glen JW (1955) The creep of polycrystalline 
ice. Proc. R. Soc. Lond. A 228:519-538

\bibitem{GreveBlatter} Greve R and Blatter H (2009) {\em 
Dynamics of Ice Sheets and Glaciers}. Springer, New York

\bibitem{Hooke} Hooke RL (2005) {\em Principles of Glacier 
Mechanics}, 2nd edn. Cambridge University Press, Cambridge, 
UK

\bibitem{MarchisottoZakeri} Marchisotto EA, Zakeri G-A 
(1994) An invitation to integration in finite terms.  
College Math. J. 25:295-308

\bibitem{Ng} Ng FSL, Barr ID, Clark CD (2010) Using the surface 
profiles of modern ice masses to inform palaeo-glacier reconstruction. 
Quat. Sci. Rev.    29:3240-3255. doi:~10.1016/j.quascirev.2010.06.045

\bibitem{Nye51a} Nye JF (1951a) The flow of glaciers and 
ice-sheets as a problem in plasticity. Proc. R. Soc. Lond. 
A 207:554-572

\bibitem{Nye51b} Nye JF (1951b) A method of calculating the 
thicknesses of  the ice-sheets. Nature 169:529-530

\bibitem{Patersonpaper} Paterson WSB (1972) Laurentide 
ice sheet: estimated volume during late Wisconsin. 
Rev. Geophys. Space Phys. 10:885-917

\bibitem{Paterson} Paterson WSB (1994) {\em The Physics of 
Glaciers}, 3rd edn. Butterworth-Heinemann, Oxford

\bibitem{Thorp} Thorp PW (1991) Surface profiles and basal shear 
stresses of outlet glaciers from a late-glacial mountain ice field in 
Western Scotland. J. Glaciol. 37:77-88

\bibitem{Vialov} Vialov SS (1958) Regularization of glacial 
shields movement and the theory of plastic viscous flow. In 
IAHS Publ. 47 (Symposium at Chamonix 1958, Physics of the 
Movement of the Ice). IAHS Press, Wallingford, UK

\bibitem{Weertman} Weertman J (1961) Stability of 
ice-age ice sheets. J. Geophys. Res. 66:3783-3792

\end{thebibliography}
\end{document}